# COOJA Network Simulator: Exploring the Infinite Possible Ways to Compute the Performance Metrics of IOT Based Smart Devices to Understand the Working of IOT Based Compression & Routing Protocols

Tayyab Mehmood

Dept. of Electrical Engineering, SEECS, NUST Islamabad

*Abstract*—this paper demonstrates the scheme regarding Internet of Things (IOT) which is well thought-out the next generation of Internet. IOT explicitly elaborates the assimilation of human beings and physical systems, as they can cooperate with each other so leading towards a sort of encroachment in networking by interconnecting things together while making use of wireless embedded systems, said to be the building blocks of IOT, that are capable to be given an IP address and thus making them part of the global internet. Several essential approaches that entail in IOT and supports this innovation are being argued in this paper. 6LoWPAN (IPV6 Low Power Personal Area Networks) is a protocol used to appropriately and efficiently use IPV6 addresses. Control messages of RPL routing protocol for low power devices are discussed to understand the working of RPL protocol. In the end Contiki OS based COOJA Network simulator is used to demonstrate the working of how these routing and compression protocol works in real time simulation.

*Keywords*—IOT, 6LoWPAN, IPV6, RPL Routing Protocol, Contiki OS, COOJA Network Simulator.

## I. INTRODUCTION

Since few decades the global internet has amended significantly in numerous ways which has put immense impact on human society as well. Advancement of Internet is still on and time to time various new innovations are emerging. This paper's core emphasis is on an innovative contraption in networking named as Internet of Things (IOT), it is one of those innovations which is going to enclose a noteworthy giving in the field of networking. IOT is "*connecting things together*" thus IOT involves interconnection of IP enabled devices called smart objects being closely related to M2M communication, and because those common things become so smart that they can comprehend their surroundings. since physical bodies are existing roughly at all places in the world these things can sense inputs and then transform them into a data set which is then processed on internet, this can be named as sensing part of IOT, inversely a certain thing can receive some data from internet and can tell you about something using actuators this can be named as actuating part of IOT. Finally summarizing IOT it can be said that the combination of physical objects plus actuators, sensors, controller plus internet formulates IOT [1].

IPV6 is the most dependable factor which is the key to the success of IOT, as there will be trillions of objects that need to be given IPs so IPV6 formulates that incredibly much promising to allocate those addresses. As there will be trillions of small objects joining the internet most of which is going to happen through wireless technology hence it is going to be a very much new internet which is going to be very much large then the current internet itself one day. The distinctive smart object (thing) implemented on IOT was a coke machine, connected to the internet, in 1980s at the Carnegie Melon University.

The union of IPSO (IP for Smart Objects) has made a prodigious work to sponsor the use of IP address for low power devices [2]. IPSO supports the usage of the layered IP architecture for low power wireless sensor nodes which have a small CPU inside them and it is also the leading association for outlining the IOT (*Internet of Things*). The collaboration of IPSO with IETF (Internet engineering task force) quickens the implementation of IPv6 on LLNs (low power and lossy networks). 6LoWPAN (IPv6 over Low power Wireless Personal Area Networks) standard gives the idea of using Ipv6 for every small and low power device and it is specified by IETF organization [3]. The 6LoWPAN standard permits the use of web services without any application gateways, because of this nodes with limited hardware and computing resources are become capable to contribute in IOT.

This paper shows the RPL (IPv6 based Routing Protocol for Low power and Lossy Networks) [4], which has been made by the IETF group to reduce the routing issues in LLNs. RPL gears up methods to cut the energy consumption of the IOT smart devices by dynamically control the sending rate of control packets and addressing topology in dependability only when data packets have to be sent. RPL use IPv6 addressing protocol and supports traffic flow not only in the downward direction (from the gateway/router to leafs), but also in the upward direction. [5] and [6] shows the effect of lossy radio links on the maximum achievable throughput, overall reliability of the link and power efficiency of the system. Due to broadcasting and retransmissions of the packets lossy radio links increases the power consumption and with lossy radio networks we could achieve roughly half of the data rate as compared to the lossless networks. [7] Shows the impact of lossy radio links and resulted that the 50% to 80% of the energy is wasted due to environmental effects in outdoor and in indoor situations and to overcome the packet collisions.

To encounter the challenges and requirements of lossy radio networks, IETF ROLL WorkingGroup designed a new routing protocol for wireless and wired networks named as RPL [8].The main aim of RPL is to offer efficient routing paths for MP2P and P2MPtraffic designs in LLNs but it is ideal for data sink communication (n-to1 or MP2P). RPL also supports the new Internet Protocol IPv6.

In the later sections of the paper are organized as follows in section II wireless sensor networks have been briefly discussed, in section III 6LoWPAN has been highlighted, section IV emphasizes on LRP, section V belongs to COOJA simulator and discussion will finally be concluded in section VI.

## II. Wireless Sensor Network

WSN is a collection of number of sensors that manipulates the physical or environmental conditions. WSNs usually possess many distributed sensors which consists of a transducer "a device used for transformation of energy from one form to another", power source, a microprocessor and a transceiver these in combination examines the condition of a physical body and then forwards the data towards a certain data center to which the network is connected. WSNs typically varies in size as some sensors have big power sources like battery and some have small power sources. WSN is a part of IOT as is based upon these sensors everything that belongs to IOT has to be coupled with such sensors so that these sensors can monitor their behavior or condition. [9] several issues that are associated with WSNs include battery timing because to have a constant connection to the internet a node using certain protocols must be on regular basis but the advantages a WSN can provide cannot be exempted, which can play a remarkable role in the progression of IOT[10].

## III. 6LoWPAN

6LoWPAN is a standard protocol that is imperative for the efficient usage of IPV6 over low power low rate wireless sensor networks via an adaption layer [11], therefore the name 6LoWPAN has been derived by the combination of IPV6 and low power sensor networks [12]. IPV6 being very heavy for low power wireless sensor networks is compressed with the help of 6LoWPAN using a range of compressing techniques so that it can be used within those low power environments efficiently. As a good number of the WSNs are being powered by batteries so energy efficiency becomes a crucial need which should be pleasing. The IEEE 802.15.4 standard offers the solution by introducing data rates of 20 to 250 kb/s depending on which frequency channel is being used [13]. During transmissions certain protocols lay much overhead, in particular, IPV6 which has much longer headers that can occupy rest of the available bandwidth which turns out to be a real setback while imposing IP to WSNs. To resolve this problem 6LoWPAN was launched in which, with the help of an adaption layer, header compression and fragmentation are made possible making the IPV6 overhead lesser and lesser.

### i. IEEE 802.15.4 and 6LoWPAN LAYRING.

Figure 1 give a picture of how a 6LoWPAN adaption layer is used to permit IPV6 over IEEE 802.15.4 standard, different layers and their respective contributions are as follows. Application layer is accountable for web communication among nodes and it makes use of HTTP protocol.-Socket presents interface between Application layer and Transport layer.-TCP/UDP protocols are used for caring Application layer messages over transport layer.IPV6 and ICMPV6 are representing network layer where IPV6 performs node to node delivery and ICMPV6 is liable for correcting errors and few more basic functionalities.6LoWPAN adaption layer takes account of header compression in order to overcome the overhead of IPV6.Media Access Control offers access to the physical channel.-Physical layer in IEEE 802.15.4 standard is used for number of functions it provides energy management, RF transceiver, and channel selection [14].

### ii. COMPRESSION TECHNIQUES

Header compression is one of the most prominent contributions of 6LoWPAN in IOT. Generally two species of compressions are encountered by 6LoWPAN stateful and stateless. In stateful compression there are certain fields who don't modify their values throughout communication whereas is stateless compression certain fields draw on common values due to which the header length shrinks [15]. 6LoWPAN presents stateless compression which is also known as HC, this technique can switch a 40 or 60 bytes into few bytes as pointed up in figure 2. The version field in IPV6 packet format can be neglected provided all nodes are using the same IPV6 version, similarly length of the packet can also be overlooked because that information can be achieved by MAC header[16], further most of the bytes are occupied by the

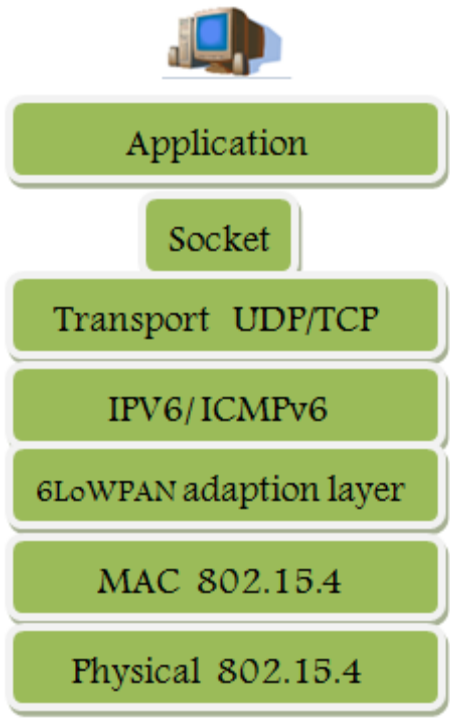

Figure 1: IEEE 802.15.4 and 6LoWPAN layering [8].

Source and destination addresses they can be compressed in three unlike ways which are classified as partly and fully compressed headers. In two partly compressed headers source and destination fields are compressed separately in two packets and in fully compressed headers both source and destination fields are compressed at a time in a single packet. Partly compressions are used among gateways and sensor nodes whereas fully compressions are used among sensor nodes within the same network [17].

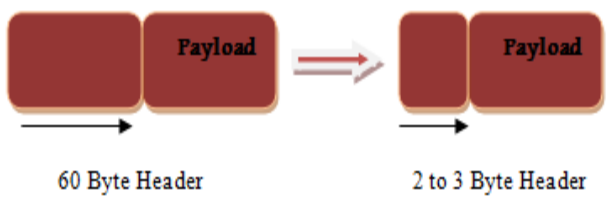

Figure 2 IP Header Compression from 40 bytes to 2-4 bytes [5].

## IV. OVERVIEW OF RPL PROTOCOL

RPL is a distance vector routing protocol for low power small devices and for lossy networks that use IPv6. These small network devices are connected in such a way that they don't create any loop. Because of this Destination Oriented Directed Acyclic Graph (DODAG) is built to route at a single terminus. This destination node is named as DODAG root in the RPL specification. The network graph is assembled by using the OF (Objective Function) which explains how the routing metric is calculated. During topology creation OF states that how routing constraints and other functions are taken into account. LLN radio links are very instable, have very low data rates and have very high loss rates and these limitations are considered while defining the OF (minimizing latency, energy, etc).

Network has to be optimized according to the different application scenarios. For instance, a DODAG may be constructed according to the battery power consumption of the device, memory of the device, processing capability of the device or according to the ETX (Expected Number of Transmissions). So network optimization is done by the RPL Instance which permits to build a logical routing topology over the present WSN arrangement and RPL protocol also states the objective function for a group of one or more DODAGs.

To avoid routing loops with respect to the DODAG root, nodes calculate its position relative to the other nodes and this position is named as Rank or Cost. If the node is mobile then Rank increases (if node move away from the DODAG root) and decreases (if the node move the direction of the DODAG root). In the network graph, DODAG root is grounded when it satisfy all the goals and it is floating when it does not specify all the goals but only in DODAG communication. Towards root, immediate successor of the leaf (child) is known as parent. These leafs are of two kinds either they store the routing tables for sub-DODAG or they do not store any routing table but only know their parents.

For information exchange and topology maintenance, RPL protocol uses four types of control messages; DODAG Information Object (DIO), Destination Advertisement Object (DAO), DODAG Information Solicitation (DIS) and DAO-ACK. Like beacon DIO multicasts the RPL instance in the downward direction to allow the other sensor nodes to stores the information about IPv6 address of the root, current RPLInstance, current rank of the node and joined it. If leaf don't hear ant announcement then it can send a request named as DIS. DIS make feasible for the leaf to request for the DIO message (neighbor discovery). DAO is the request from the leaf to the root/parent to join it on the DODAG as a child. DAO-ACK is a response of DAO message which is sent by the root or patent (root recipient) to the leaf. In RPL network there are three kinds of nodes. The first one is root nodes which offer the connectivity to the leaf nodes and sometimes it is also known as gateway node. The second one is router which is used to advertise the topology information and routing tables to the neighbors. The third one is leaf (child node) which has the ability to join the DODAG and it does not have the ability to send DIO message.

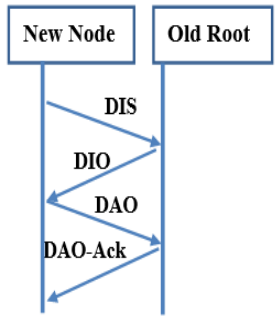

Fig 3: Flow diagram of control messages of RPL protocol

Most of the routing protocols usually broadcast the control messages at a constant rate which cause the waste of energy of the device when it is in a stable condition. Therefore RPL protocol uses the Trickle algorithm to control the sending rate of DIO messages [10]. The control messages will be exceptional in a network with stable links however control messages will send more often in the situation where the topology changes repeatedly.

### 1. STRUCTURE OF DIO CONTROL MESSAGE

To construct the topology, DIO message is the main source of information. Figure 2 shows the structure of the DIO message. After sending the DIS request message, node receive the DIO and discover the RPL instance by storing the first data field of the DIO control message.

| PRLInstanceID | | | | Version Number | RANK | |
|---|---|---|---|---|---|---|
| G | O | MOP | Prf | DTSN | Flags | Reserved |
| DODAGID (128 bit) | | | | | | |

Fig 4: Structure of DIO control message

RPL instance ID is used to uniquely identify the set of independent DODAGs best for the given situation. 128 bit DODAG ID is the routable IPv6 address of the root node. Whenever DODAG reconstruction is needed, DODAG version number is incremented. "G" tells whether the DODAG is grounded or not. "MOP" (mode of Operation) is arranged by the DODAG root and it is used for downward routing. "Prf" field defines the preferable root node and it is of 3 bits. DSTN is used to store the sequence number and is used by the node to see the freshness of the DIO message.

### 2. ROUTING LOOPS IN LLNs:

Because of link failure or mobility network topology may change and a node may take a different path for the given destination. Loop may occur if the child node picked as the next hop. Because of this loop, delays, waste of bandwidth and device energy, packet drops and network congestion may occur in LLNs. Therefore, RPL protocol for LLNs must explain a loop avoidance mechanism during the topology construction.

#### i. LOOP AVOIDENCE STRATEGY

Up till now, in the network graph Rank is responsible for the node's position and router multicast the DIO messages to the neighboring nodes for the topology maintenance. If every node in the range accept the multicast message and consider the DIO message sender for the calculation of parent set then it may be possible that child nodes are chosen as the next hope.

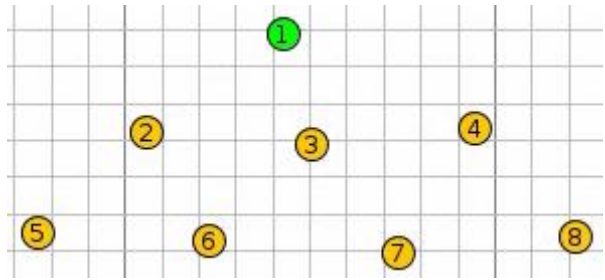

Fig 5: Sink(green) and sources (yellow) in the COOJA Network Simulator Enviornment

If the node 2 computes the DIO message from the node 5 and consider it as a effective potential parent. If the link between node 2 and root node 1 fails then node 2 will automatically consider its child node (5) as the next hop and eventually loop will occur. Since ETX cost via node 5 to the root node 1 is 3 and ETX cost via node 2 to the root node is 2 therefore node does not analyse the DIO message from the higher Rank nodes than itself.

### ii. RPL METRICS IN LLNs:

Most of the routing protocols only consider the link metrics of the network and don't consider the node's current status but this may be critical for the LLNs where nodes have limited resources and are battery powered. Node's status comprises of available memory, remaining energy of the battery and the CPU usage. For instance, if a chain topology happens in the WSN deployment then the last node before the root node will typically experience more forwarding overhead and greater traffic load. In figure 5 all the nodes in the network produce some data packets and send them to the root node, in this manner node 2 may fail rapidly because of the extra energy consumption. Although link between node 3 and node 6 is not the best path but it may be sensible to send the data via node 6 because it offers a more stable node condition. As a result, in [7] numerous metric categories are considered before choosing the next hop by ROLL Working Group.

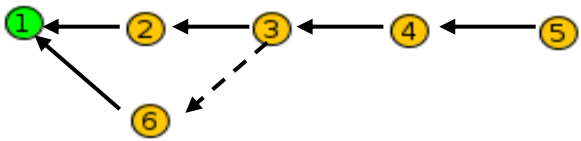

Fig 5: Chain Topology in LLNs

### iii. ENERGY CONSUMPTION OF IN LLNs:

Energy consumption of nodes method suggest that before choosing the next hop as a possible parent a node should weigh up the energy level of its neighbors. RPL metric specification use two fields of data. Type field define the type of the node, either the node is on batteries or on power. If the network device is powered it means that it may be a data collector or a root node. During parent selection or next hop selection such powered nodes are preferable. Second field includes the EE (Energy Estimation). Network devices on batteries must calculate the EE value before selecting the parent. EE is the ratio of the Power-now (remaining energy) to the Power-max (estimated power at boot up).

### iv. EXPECTED NUMBER OF TRANSMISSIONS IN LLNs:

Expected number of transmissions to send the data messages to the root node. The value of ETX is 1 for the network device which is one hop away from the root node and have strong signal strength with lossless path model. ETX calculates the link quality of a single hop between two neighboring nodes. PRR (packet reception rate) is used to calculate the link quality and it is computed at the receiver node. PRR is the ratio of the number of the received data packets to the number of send data packets.

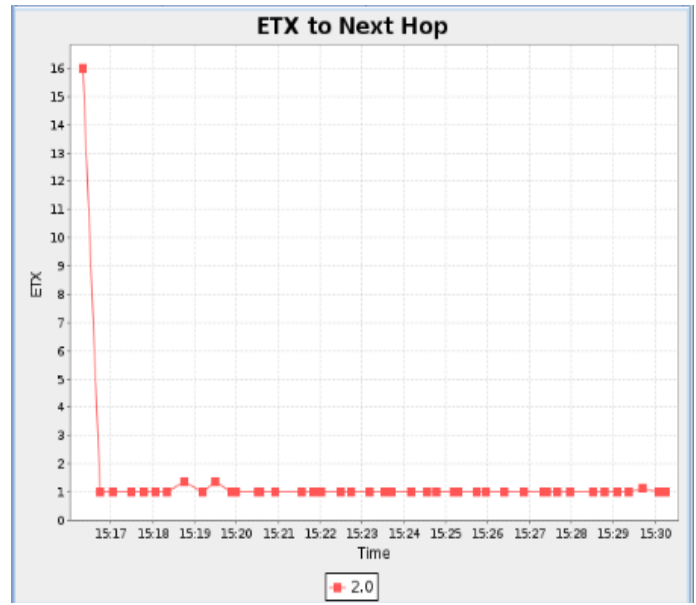

Fig 6: expected number of transmissions required to send the data to the data collector node

### V. COOJA NETWORK SIMULATOR OVERVIEW

COOJA is a network simulator which permits the emulation of real hardware platforms. COOJA is the application of Contiki OS concentrating on network behavior. COOJA is capable of simulating wireless sensor network without any particular mote. Cooja supported following set of standards; TR 1100, TI CC2420, Contiki-RPL, IEEE 802.15.4, uIPv6 stack and uIPv4 stack. There are four propagation models in the COOJA simulator which must be selected before starting a new simulation [16]. The first model is constant loss Unit Disk Graph Medium (UDGM) and it take the ideal transmission range disk in which motes inside the transmission disk receive data packets and motes outside the transmission disk do not get any packet. The second model is distance loss UDGM is the extension of constant loss UDGM and it also consider the radio interferences. Packets are transmitted with "success ratio TX" probability and packets are received with probability of "success ratio RX". The third model is Directed Graph Radio Medium (DGRM) and it states the propagation delays for the radio links. Last path loss model is multipath Ray-tracer Medium (MRM) and it uses the ray tracing methods such as Friis formula to calculate the receiver power. MRM is also capable of computing the diffractions, reflections and refractions along the radio links [17].

### i. COOJA SIMULATION INTERFACE

COOJA network simulator interface comprises of five windows. The network window displays the physical arrangement of the motes. In order to build a topology, one could change the physical position of the motes. In network window, all the different have different colors according to their functionality, i.e. sink mote has a green color and the sender mote has the yellow color. Mote attributes, radio environment of each mote, mote type and radio traffic between the motes could also the seen visually in the network windows. Simulation control window helps us to control the speed of the simulation and to pause, start and reload the current running simulation. Note window is used to write the theory and key points of the simulation and save them in the note window.

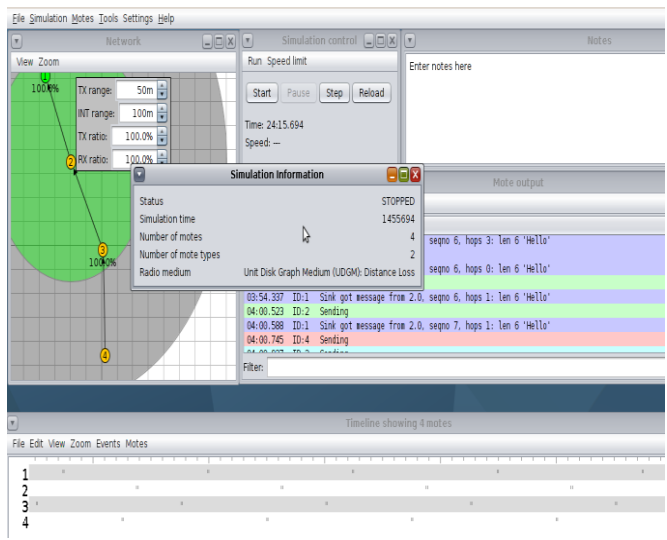

Fig 7: different windows of COOJA network simulator to compute the performance matrics

Cooja network simulator shows a timeline for each mote in the running simulation. We could use timeline for visualizing the both the power consumption and network traffic in the wireless sensor networks. In row three for mote 1, Color of the mote shows the power state of the transceiver: if the mote is off then it is white, on then it is gray as shown for mote 1. White and gray color is either hardware is off or on but the red color line in the second row shows that whenever the node hardware goes on its radio transceiver is also goes on. In first row in timeline of mote 1, Radio transmissions are shown by blue color, reception by green and radio interference is shown by red.

## VI. CONCLUSION

Change in the world of communication among several entities have been a burning issue for a long time and has showed a great progress in the form of numerous innovations among which IOT is of great significance which is all about making communication, possible, among different devices, using sensors and actuators. This paper formulates working of compression and routing protocols in IOT environment. In this paper, we see how COOJA network simulator enables the emulation different kinds of motes and how the routing matrices are computed. COOJA simulator is also used as a power visualizer in this paper.

## REFRENCES

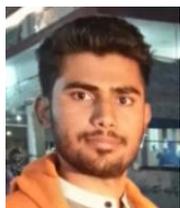
Tayyab Mehmood is a Post Graduate student of EE in SEECS, NUST Islamabad. He did his bachelor from the Islamia University Bahawalpur. He is the author of several papers which were published in reputed journals of Elsevier & in IEEE Conferences.